\documentclass[apj]{emulateapj} \slugcomment{Submitted to the
Astrophysical Journal on June 09, 2008}
\usepackage{graphicx,amssymb,amsmath,times}

\newcommand{\lsi}{LS~I~+61$^{\circ}$303}
\newcommand{\lomb}{Lomb-Scargle}
\newcommand{\eg}{3EG~J0241$+$6103}

\begin{document}
\title{Periodic very high energy $\gamma$-ray emission from \lsi\ observed with the MAGIC telescope}
\shorttitle{Periodic VHE $\gamma$-ray emission from \lsi\ }
\shortauthors{J. Albert et~al.}

%
\author{
J.~Albert\altaffilmark{a},
E.~Aliu\altaffilmark{b},
H.~Anderhub\altaffilmark{c},
L.~A.~Antonelli\altaffilmark{d},
P.~Antoranz\altaffilmark{e},
M.~Backes\altaffilmark{f},
C.~Baixeras\altaffilmark{g},
J.~A.~Barrio\altaffilmark{e},
H.~Bartko\altaffilmark{h},
D.~Bastieri\altaffilmark{i}
% ...
}

\altaffiltext{a} {Universit\"at W\"urzburg, D-97074 W\"urzburg, Germany}
\altaffiltext{b} {IFAE, Edifici Cn., Campus UAB, E-08193 Bellaterra, Spain}
\altaffiltext{c} {ETH Zurich, CH-8093 Switzerland}
\altaffiltext{d} {INAF National Institute for Astrophysics, I-00136 Rome, Italy}
\altaffiltext{e} {Universidad Complutense, E-28040 Madrid, Spain}
\altaffiltext{f} {Technische Universit\"at Dortmund, D-44221 Dortmund, Germany}
\altaffiltext{g} {Universitat Aut\`onoma de Barcelona, E-08193 Bellaterra, Spain}
\altaffiltext{h} {Max-Planck-Institut f\"ur Physik, D-80805 M\"unchen, Germany}
\altaffiltext{i} {Universit\`a di Padova and INFN, I-35131 Padova, Italy}
% ...

\begin{abstract}

The MAGIC collaboration has recently reported the discovery of
$\gamma$-ray emission from the binary system \lsi\ in the TeV energy
region. Here we present new observational results on this source
in the energy range between $300\textrm{ GeV}$ and
$3\textrm{ TeV}$. In total 112 hours of data were taken between
September and December 2006 covering 4 orbital cycles of this
object. This large amount of data allowed us to produce an integral
flux light curve covering for the first time all orbital phases of \lsi.
In addition, we also obtained a differential energy spectrum for two
orbital phase bins covering the phase range $0.5 < \phi<0.6$ and $0.6 <
\phi<0.7$.
The photon index in the two phase bins is consistent within the errors
with an average index $\Gamma=2.6\pm0.2_{stat}\pm0.2_{sys}$.
\lsi\ was found to be variable at TeV energies on timescales of
days. These new MAGIC measurements allowed us to search  for
intra-night variability of the VHE emission; however, no evidence
for flux variability on timescales down to 30 minutes was found.
To test for possible periodic structures in the light curve, we
apply the formalism developed by Lomb and Scargle to the \lsi\ data
taken in 2005 and 2006. We found the \lsi\ data set to be periodic
with a period of (26.8$\pm$0.2)~days (with a post-trial chance
probability of 10$^{-7}$), close to the orbital period.

\end{abstract}


\section{Introduction}

The $\gamma$-ray binary system \lsi\ is located at a distance of
$\sim$2 kpc and is composed of a compact object of unknown nature
(neutron star or black hole) orbiting a Be star in a highly
eccentric orbit ($e=0.72\pm0.15$ or $e=0.55\pm0.05$
following~\cite{Casares:2005wn} and \cite{Grundstrom:2006}
respectively).

\lsi\ was found to display periodic variability in the radio,
infrared, optical, and X-ray bands (\citealt{1982ApJ...255..210T},
\citealt{Paredes:1995}, \citealt{Mendelson:1989} and
\citealt{Paredes:1997}, respectively).

The orbital period of the system is 26.4960 days
long~\citep{Gregory:2002}. The periastron passage, derived from the
optical spectra, is found to be at phase $\phi=0.23\pm0.02$
in~\cite{Casares:2005wn} and $\phi=0.301\pm0.011$
in~\cite{Grundstrom:2006}, adopting a zero-phase at $T_0=\textrm{JD
} 2443366.775$.

Radio outbursts are observed every orbital cycle at phases varying
between 0.45 and 0.95 with a 4.6 years
modulation~\citep{Gregory:2002}. Radio imaging techniques have shown
extended, radio-emitting structures with angular extensions of
$\sim$0.01 to $\sim$0.1 arc-sec, where the radio emission originates
in a two-sided, possibly precessing, relativistic jet
($\beta/c=0.6$)~\citep{Massi:2004}. These extended radio structures
have led some authors to adopt the microquasar scenario to explain
the non-thermal emission in \lsi\ ~\citep[e.g.,][]{Bosch-Ramon:2006,
Bednarek:2006}. Recent high resolution VLBA measurements show a
complex and changing morphology different from what is expected for
a typical microquasar jet (see radio images
in~\citealt[e.g.,][]{2006smqw.confE..52D,magiclsi.2006MW}).
Furthermore no solid evidence for the presence of an accretion disk
(i.e. a thermal X-ray component) has been
observed~\citep{2006MNRAS.372.1585C}. This seems to favor a scenario
in which the non-thermal emission in \lsi\ is powered by the
interaction between a pulsar and the primary star
winds~\citep{Maraschi:1981}.

At higher energies, \lsi\ was found to be spatially coincident with
the EGRET $\gamma$-ray source \eg~\citep{kniffen}. Variable emission at
TeV energies was observed with the MAGIC
telescope~\citep{Albert:2006vk} and was recently confirmed by
VERITAS~\citep{Veritas_lsi}. The system showed the peak TeV
$\gamma$-ray flux at phase $\phi\sim0.65$, while no very high-energy
emission was detected around the periastron passage.

Here we present new MAGIC telescope observations of \lsi. We briefly
discuss the observational technique and the data analysis procedure,
investigate the very high energy (VHE) $\gamma$-ray spectrum during
the high emission phase of the source, and put the results into
perspective with previous VHE $\gamma$-ray observations of this
system. Finally we analyzed the temporal characteristics of the TeV
emission and find a periodic modulation of the signal with the
orbital period.

\section{Observations}

The observations were performed from September to December 2006
using the MAGIC telescope on the Canary island of La Palma
($28.75^\circ$N, $17.86^\circ$W, 2225~m a.s.l.), from where \lsi\ is
observable at zenith distances above $32^\circ$. The telescope
operates in the energy band from $50-60$~GeV (trigger threshold at
zenith angles less than 30 degrees) up to tens of TeV, with a
typical energy resolution of $20-30\%$. The accuracy in
reconstructing the direction of the incoming $\gamma$-rays is about
$0.1^{\circ}$, depending on the energy. A detailed description of
the telescope performance can be found in~\cite{crab:2008}.

The data on \lsi\ were taken between 15$^{\textrm{th}}$ of September
2006 and 28$^{\textrm{th}}$ of December 2006 covering 4 orbital
periods of the system. In total 120 hours of data were taken at
zenith angles between 32 and 55$^{\circ}$, with $\sim97$\% of the
data below $44^{\circ}$. After pre-selection of good quality data a
total of 112  hours of data remained for the analysis. About 17\% of
these were recorded under moderate moonlight conditions. Due to the
different observation conditions such as bad weather, too bright
moon or too large zenith angle, the data set was not uniform with
the orbital phase. In Table~\ref{tab:LC} the observation times of
the analyzed data are summarized.

The observations were carried out in wobble
mode~\citep{Fomin:1994aj}, i.e. by alternately tracking two
positions at $0.4^\circ$ offset from the actual source position.
This observation mode allows for a reliable background estimate for
point like objects such as \lsi.

\section{Data Analysis}

The data analysis was carried out using the standard MAGIC analysis
and reconstruction software~\citep{crab:2008} . The images were
cleaned by requiring a minimum number of 10 photoelectrons (core
pixels) and 5 photoelectrons (boundary pixels), see
e.g.~\citet{Fegan:1997db}. The quality of the data was checked and
bad data such as accidental noise triggers or data taken during
adverse conditions (very low atmospheric transmission, car light
flashes etc.) were rejected. From the remaining events, image
parameters were calculated~\citep{Hillas_parameters}.

For the $\gamma$/hadron separation a multidimensional classification
procedure based on the Random Forest method~\citep{magic:RF,
  Bock:2004td} was used.
For every event a parameter called hadronness ($h$) is derived,
based on the values of the events image parameters. The hadronness
denotes the probability that an event is a hadronic induced
(background) event. The final separation was achieved by a cut in
$h$ which was determined by requiring 80\% of the simulated Monte
Carlo (MC) $\gamma$-ray events to be kept. In addition to the cut in
$h$ a geometrical cut in the squared angular distance of the assumed
source position to the shower direction axis ($\theta^2\textrm{
cut}$) was performed so that $70\%$ of all simulated MC $\gamma$-ray
events from a point-like source are left after the cut. The cut
efficiencies were determined by optimizing the significance of a
Crab Nebula data sample recorded under the same zenith angle as the
\lsi\ data set. The same cut procedure was applied to the final
\lsi\ sample. The energy of the primary $\gamma$-ray was
reconstructed from the image parameters using also a Random Forest
method leading to an assigned estimated energy for each
reconstructed $\gamma$-ray event. The differential energy spectrum
is unfolded taking into account the full instrumental energy
resolution~\citep{magic:unfolding}. For the integral flux
calculation of the light curves we used fixed cuts (for all
energies) in hadronness and $\theta^2$. In the case of the energy
spectrum determination we derived fixed hadronness and $\theta^2$
cuts for each energy bin.

The main contributions to the systematic error of our analysis are
the uncertainties in the atmospheric transmission, the reflectivity
(including stray-light losses) of the mirror and the light catchers,
the photon to photoelectron conversion calibration and the
photoelectron collection efficiency in the photomultiplier
front-end~\citep{crab:2008}. Also MC uncertainties in the detector
simulation and systematic uncertainties from the analysis methods
contribute significantly to the overall error.\\
All errors in this paper are statistical errors, otherwise it is
stated explicitly. In addition there is a $30\%$ systematic
uncertainty on flux levels and $0.2$ on the spectral photon index.

MAGIC has the capability to operate under moderate moonlight. This
permits to increase the duty cycle by up to 28\%, thus considerably
improving the sampling of transient sources. In particular, 17\% of
the data used in this analysis were recorded under moonlight. The
nights which were partly taken under moonlight conditions are
labeled with a star in Table\ref{tab:LC}. For these days we estimate
an increased systematic error of $\sim 40\%$ instead of the $\sim
30\%$ in the case of the dark night observations. All spectra are
derived from data which is only taken under dark night conditions
and thus no additional error is present in the obtained parameters.

\subsection{Light Curve}

\begin{figure}[tbp]
  \centering
  \includegraphics[width=\linewidth]{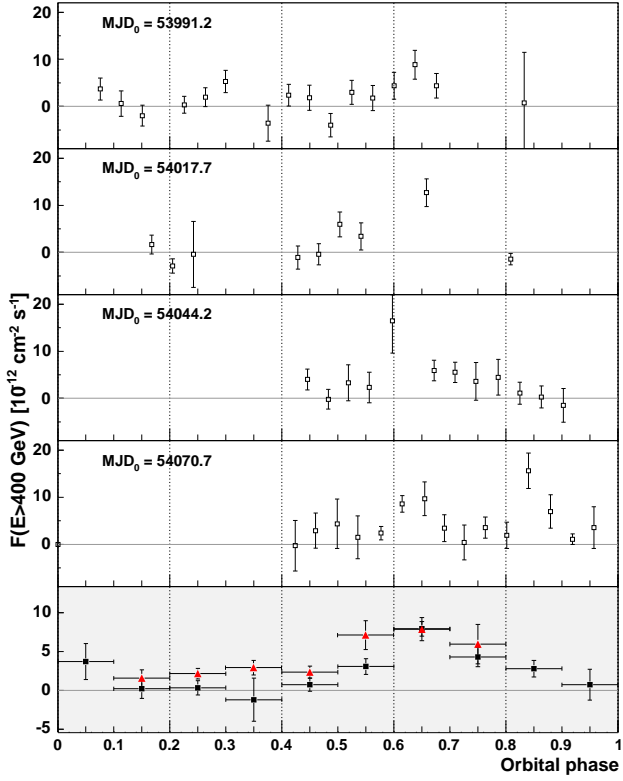}
  \caption{VHE ($E>400\text{GeV}$) gamma-ray flux of \lsi\ as a
    function of the orbital phase for the four observed orbital cycles
    (4 upper panels) and averaged for the entire observation time
    (lowermost panel). In the lower most panel the previous published
    \citep{Albert:2006vk} averaged fluxes per phasebin
     are shown in red too. Vertical error bars include $1\sigma$ statistical
    error.}
  \label{fig:lc}
\end{figure}

Figure~\ref{fig:lc} presents the gamma-ray flux above 400~GeV
measured from the direction of \lsi\ as a function of the orbital
phase of the system for the 4 observed orbital cycles. The
probability for the distribution of measured fluxes to be a
statistical fluctuation of a constant flux (obtained from a $\chi^2$
fit to the entire data sample) is $4.4\times10^{-6}$
($\chi^2/\textrm{dof} = 108.9/51$). In all orbital cycles
significant detections ($S > 2\sigma$) occurred during the orbital
phase bin 0.6--0.7. The highest measured fluxes are dominantly found
in this phase bin. Among those nights around phase 0.65, the night
MJD 54035.11 shows
the maximum flux, with statistical significance of 4.5~$\sigma$.\\

At the periastron passage (phase 0.23, according
to~\citealt{Casares:2005wn}) the flux level is always below the
MAGIC sensitivity and we derive an upper limit with $95\%$
confidence level of $4 \times 10^{-12}$~cm$^{-2}$s$^{-1}$ (MJD
53997). If we take for the periastron passage the phase value 0.3 as
obtained by~\cite{Grundstrom:2006}, we detect a marginal signal on
MJD 53999 with a flux of $F(\textrm{E}>400 \textrm{ GeV})=5.3\pm2.4
\times 10^{-12} \textrm{cm}^{-2}\textrm{s}^{-1}$. Since the correct
value for the periastron passage is yet debated we can put strong
constrains to the emission only in the case of phase 0.23.

\begin{table}[tbp]
  \caption{Observation time, orbital phase, integral flux (above
    400~GeV), flux upper limit at the 95\% confidence level (given in
    case flux significance is $\lesssim 2 \sigma$,
    ~\citep[following]{Rolke:2004mj}). Nights partly taken under moonlight conditions
    are labeled with a star.}
  \vspace*{-0.3cm}
  \begin{center}
    \begin{tabular}{lcccc}
      \hline \hline
     Middle Time    & Obs. Time & Phase & Flux & Upper limit \\
      (MJD)   & (min)     &            & 10$^{-12}$          & 10$^{-12}$            \\
      &           &            & (cm$^{-2}$ s$^{-1}$)& (cm$^{-2}$ s$^{-1}$)  \\
\hline
53993.18$^{*}$ & 137 & 0.08 & 3.7 $\pm$ 2.3 & 8.5\\
53994.17$^{*}$ & 112 & 0.11 & 0.6 $\pm$ 2.7 & 6.2\\
53995.17$^{*}$ & 157 & 0.15 & $-$2.0 $\pm$ 2.2 & 3.0\\
53997.15 & 229 & 0.23 & 0.3 $\pm$ 1.8 & 4.0\\
53998.15 & 211 & 0.26 & 2.0 $\pm$ 2.0 & 6.0\\
53999.10 & 133 & 0.30 & 5.3 $\pm$ 2.4 & \nodata\\
54001.12 & 82 & 0.38 & $-$3.6 $\pm$ 3.8 & 5.1\\
54002.09 & 188 & 0.41 & 2.4 $\pm$ 2.3 & 7.1\\
54003.08 & 144 & 0.45 & 1.8 $\pm$ 2.7 & 7.2\\
54004.08 & 158 & 0.49 & $-$4.0 $\pm$ 2.5 & 2.5\\
54005.07 & 155 & 0.52 & 3.0 $\pm$ 2.5 & 8.1\\
54006.07 & 162 & 0.56 & 1.8 $\pm$ 2.7 & 7.2\\
54007.08 & 139 & 0.60 & 4.4 $\pm$ 2.8 & 10.2\\
54008.07 & 152 & 0.64 & 8.8 $\pm$ 3.1 & \nodata\\
54009.08 & 147 & 0.68 & 4.4 $\pm$ 2.6 & 9.7\\
54013.24 & 7 & 0.83 & 0.8 $\pm$ 10.7 & 26.7\\
\hline
54022.10$^{*}$ & 186 & 0.17 & 1.7 $\pm$ 2.0 & 5.8\\
54023.10$^{*}$ & 269 & 0.20 & $-$2.9 $\pm$ 1.5 & 1.4\\
54024.08$^{*}$ & 20 & 0.24 & $-$0.4 $\pm$ 7.0 & 15.2\\
54029.02 & 134 & 0.43 & $-$1.1 $\pm$ 2.5 & 4.1\\
54030.01 & 161 & 0.47 & $-$0.4 $\pm$ 2.3 & 4.2\\
54031.01 & 163 & 0.50 & 5.9 $\pm$ 2.6 & \nodata\\
54032.01 & 139 & 0.54 & 3.4 $\pm$ 2.9 & 9.2\\
54035.11 & 150 & 0.66 & 12.7 $\pm$ 2.9 & \nodata\\
54039.09 & 93 & 0.81 & $-$1.4 $\pm$ 1.2 & 1.7\\
\hline
54055.97 & 181 & 0.45 & 4.0 $\pm$ 2.2 & 8.5\\
54056.96 & 223 & 0.48 & $-$0.2 $\pm$ 2.1 & 4.2\\
54057.90 & 66 & 0.52 & 3.3 $\pm$ 3.8 & 11.2\\
54058.90 & 57 & 0.56 & 2.3 $\pm$ 3.3 & 9.2\\
54060.00 & 17 & 0.60 & 16.5 $\pm$ 6.8 & \nodata\\
54061.96 & 221 & 0.67 & 5.9 $\pm$ 2.2 & \nodata\\
54062.96 & 228 & 0.71 & 5.5 $\pm$ 2.1 & \nodata\\
54063.95 & 56 & 0.75 & 3.6 $\pm$ 4.0 & 12.1\\
54065.00$^{*}$ & 71 & 0.79 & 4.5 $\pm$ 3.8 & 12.4\\
54066.02$^{*}$ & 185 & 0.82 & 1.1 $\pm$ 2.4 & 5.9\\
54067.04$^{*}$ & 188 & 0.86 & 0.3 $\pm$ 2.3 & 5.0\\
54068.08$^{*}$ & 77 & 0.90 & $-$1.5 $\pm$ 3.6 & 6.1\\
\hline
54081.89 & 17 & 0.42 & $-$0.3 $\pm$ 5.4 & 12.6\\
54082.85 & 77 & 0.46 & 2.9 $\pm$ 3.7 & 10.6\\
54083.88 & 31 & 0.50 & 4.4 $\pm$ 5.3 & 15.9\\
54084.85 & 63 & 0.54 & 1.5 $\pm$ 4.5 & 10.8\\
54085.95 & 111 & 0.58 & 2.4 $\pm$ 1.4 & 5.5\\
54086.95 & 282 & 0.61 & 8.6 $\pm$ 1.8 & \nodata\\
54088.01 & 82 & 0.65 & 9.7 $\pm$ 3.6 & \nodata\\
54088.95 & 83 & 0.69 & 3.4 $\pm$ 2.9 & 9.4\\
54089.89 & 29 & 0.73 & 0.4 $\pm$ 3.7 & 9.0\\
54090.88 & 176 & 0.76 & 3.6 $\pm$ 2.2 & 8.1\\
54091.90 & 140 & 0.80 & 1.9 $\pm$ 2.8 & 7.6\\
54092.92 & 92 & 0.84 & 15.6 $\pm$ 3.8 & \nodata\\
54093.97 & 92 & 0.88 & 7.0 $\pm$ 3.5 & \nodata\\
54095.01$^{*}$ & 57 & 0.92 & 1.1 $\pm$ 1.1 & 4.1\\
54096.02$^{*}$ & 49 & 0.96 & 3.6 $\pm$ 4.4 & 12.8\\
\hline
    \end{tabular}
  \end{center}
  \label{tab:LC}
\end{table}

Summing up all data between phase 0.6 and 0.7, where the maximum
flux level is observed, we determine an integral flux above 400 GeV
of
\begin{equation*}
F(E>400 \mathrm{GeV}) =  (7.9 \pm 0.9_\text{stat} \pm
2.4_\text{syst}) \times 10^{-12} \mathrm{cm}^{-2} \mathrm{s}^{-1}.
\end{equation*}

The above quoted flux corresponds to $7\%$ of the integral Crab
nebula flux in the same energy range. The mean flux for all other
phase bins can be found in Table~\ref{tab:LCmean}.
This is well in agreement with the flux measured by MAGIC in the first
campaign~\citep{Albert:2006vk}. The data we presented here have been
reanalized with an improved energy estimation.

\begin{table}[tbp]
\caption{Average flux level above 400~GeV for each orbital phase
    bin. Flux upper limit at the 95\% confidence level are quoted in
    case flux significance is $\lesssim 2 \sigma$
    (following~\cite{Rolke:2004mj}). }
\vspace*{-0.3cm}
\begin{center}
\begin{tabular}{ccc}
\hline \hline
 Phase bin & Flux & Upper Limit \\
           & (10$^{-12}$cm$^{-2}$ s$^{-1}$)& (10$^{-12}$cm$^{-2}$ s$^{-1}$)\\
\hline
0.0--0.1         & 3.7  $\pm$ 2.3   & 8.5 \\
0.1--0.2         & 0.2  $\pm$ 1.2   & 2.7   \\
0.2--0.3         & 0.3 $\pm$ 0.9   & 2.2    \\
0.3--0.4         & -1.2 $\pm$ 2.8   & 4.3   \\
0.4--0.5         & 0.7  $\pm$ 0.8   & 2.4    \\
0.5--0.6         & 3.1  $\pm$ 1.0   & \nodata \\
0.6--0.7         & 7.9  $\pm$ 0.9   & \nodata \\
0.7--0.8         & 4.3  $\pm$ 1.2   & \nodata \\
0.8--0.9         & 2.8  $\pm$ 1.1   & \nodata \\
0.9--1.0         & 0.7 $\pm$ 2.0   & 4.8   \\
\hline
\end{tabular}
\end{center}
\label{tab:LCmean}
\end{table}

One very interesting peculiarity found in the light curve is that a
second peak in the flux level is seen in the last observed period at
phase 0.84. The flux is $(16\pm4) \times 10^{-12}$~cm$^{-2}$s$^{-1}$
which is at a similar level compared to the maximum flux detected in
phase $\sim$0.65 (MJD 54035.11). This high flux was not seen in
similar phases in any previous cycle, where only upper limits could
be set (see Table~\ref{tab:LC}). Six hours after our measurement,
the data of the Swift X-ray satellite showed a high flux
($0.25\pm0.01 \textrm{ counts s}^{-1}$)at phase
0.85~\citep{2007A&A...474..575E}. These Swift observations did not
cover the same orbital phase in any other orbit. Beside this second
peak the main X-ray emission peak is found between the phases
0.5--0.8 ($\sim 0.24\textrm{ counts s}^{-1}$
~\citealt{2007A&A...474..575E}) in exactly the same phase bin where
\lsi\ is detected by MAGIC. This is a hint for a correlated
X-ray/TeV emission.

Our measurement is in agreement with the published VERITAS
measurements~\citep{Veritas_lsi}, that \lsi\ is detected at TeV
Energies in the phase range 0.5--0.8. The MAGIC and VERITAS data are
not strictly simultaneous taken and VERITAS did not observe \lsi\ in
December~2006 were the second peak occurred.\\
Due to our long observation time and dense sampling of orbital
phases we obtained the currently most detailed light curve of \lsi .

\subsection{Spectral studies}

As seen from the light curve (Fig.~\ref{fig:lc}) \lsi\ is a very
variable source which shows high flux levels only at some orbital
phases. For the phases $0.5 < \phi < 0.6$ and $0.6 < \phi < 0.7$,
where we measured significantly high flux levels we were able to
determine differential energy spectra. In both cases the obtained
energy spectra are compatible with pure power laws. In the case of
the phase bin $0.6 < \phi < 0.7$ a power law fit gives:
\begin{equation*}
\frac{\mathrm{d}F}{\mathrm{d}E} =
\frac{(2.6\pm0.3_\text{stat}\pm0.8_\text{syst}) \cdot 10^{-12}}
{\mathrm{TeV}\,\mathrm{cm}^2\,\mathrm{s}} \left(\frac{E}{1\,
\mathrm{TeV}}\right)^{-2.6\pm0.2_\text{stat}\pm0.2_\text{syst}} \, ,
\end{equation*}
with a reduced $\chi^2/\textrm{dof} = 5.22/5$. The spectral fit
parameters agree excellent with the previous reported ones by
MAGIC~\citep{Albert:2006vk}.

In addition we derived the differential energy spectra for the two
nights with a signal of~$ > 4.5\sigma$ significance, which are part
of the same phase bin $0.6 < \phi < 0.7$. Both spectra are also well
described by a pure power law (see table~\ref{tab:spectra}). No
evidence for spectral variations has been found.

In case of phase bin $0.5 < \phi < 0.6$ we obtained:
\begin{equation*}
\frac{\mathrm{d}F}{\mathrm{d}E} =
\frac{(1.2\pm0.4_\text{stat}\pm0.3_\text{syst}) \cdot 10^{-12}}
{\mathrm{TeV}\,\mathrm{cm}^2\,\mathrm{s}} \left(\frac{E}{1\,
\mathrm{TeV}}\right)^{-2.7\pm0.4_\text{stat}\pm0.2_\text{syst}} \, ,
\end{equation*}
with a reduced $\chi^2/\textrm{dof}=1.42/4$, showing that the
spectral shape is well compatible with a simple power law.

\begin{figure}[tbp]
  \centering
  \includegraphics[width=\linewidth]{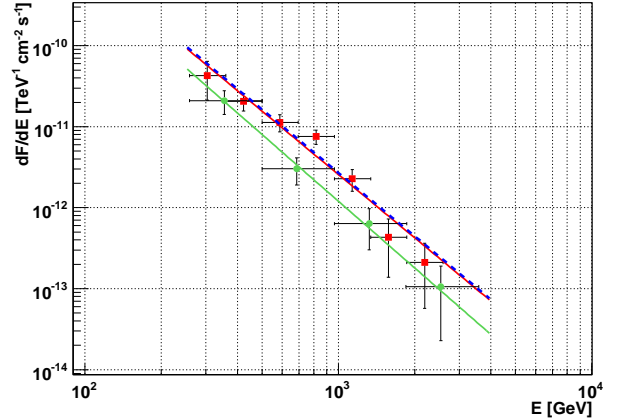}
  \caption{Shown is the spectrum of phase bin $0.5 < \phi < 0.6$
    (green), phase bin $0.6 < \phi < 0.7$ (red) and the spectrum
    obtained from previous MAGIC measurements (blue
    dashed)~\cite{Albert:2006vk}. The spectra from phase bin $0.6 < \phi
    < 0.7$ and the previous MAGIC measurements can be well described by
    a simple power law with photon index $2.6\pm0.2$. The
    spectral slope of phase bin $0.5 < \phi < 0.6$ is compatible with these
    results within the errors.}
  \label{fig:spec05}
\end{figure}

The energy spectra of the two phase bins together with the power law
fits are shown in Fig.~\ref{fig:spec05}. The corresponding fit
parameters and their errors are also shown in
Table~\ref{tab:spectra}.

\begin{table}[tbp]
\caption{Spectral fitting parameters} \vspace*{-0.3cm}
\begin{center}
\begin{tabular}{ccc}
\hline \hline
 Phase / MJD & Flux & Spectral Photon Index \\
           & (10$^{-12} $TeV$^{-1}$cm$^{-2}$ s$^{-1}$)&  \\
\hline
$0.6-0.7$ & $2.6\pm0.3$ & $2.6\pm0.2$\\
$0.5-0.6$ & $1.2 \pm 0.4$ &$2.7 \pm 0.4$\\
\hline
54035 &$3.6\pm1.1$ & $2.7\pm0.5$\\
54086 &$3.2\pm0.6$ & $2.6\pm0.3$\\
 \hline
\end{tabular}
\end{center}
\label{tab:spectra}
\end{table}

The spectral indices of the fitted power laws, for both phase bins
and the single night spectra, are compatible within their errors
indicating that no significant spectral changes happened between the
different phase bins and between the different orbital cycles. In
the phase bins $0.0 < \phi < 0.5$ and $0.7 < \phi < 1.0$ the
$\gamma$-ray flux is too low to derive meaningful differential
energy spectra.

\begin{figure}[tbp]
  \centering
  \includegraphics[width=\linewidth]{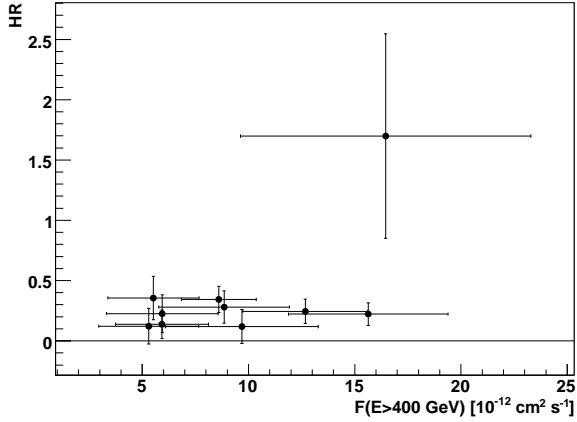}
  \caption{The $HR$, defined as $F(E>900\textrm{GeV}) /F(400<E<900\textrm{ GeV})$, vs. the integral flux
$F(E>400\textrm{ GeV})$. Each point
    is one night with a reconstructed signal of at least
    $2\sigma$~significance. There is no clear correlation
    between the $HR$ and the flux level.}
  \label{fig:HR}
\end{figure}

Another possibility to search for spectral variation is by means of
the hardness ratio $HR$ which we define as the ratio of the integral
flux between 400~GeV and 900~GeV and  above 900~GeV. The $HR$
plotted against the total integral flux above 400~GeV for each night
with a signal above $2\sigma$ significance is shown in
Fig.~\ref{fig:HR}. The requirement of the $2\sigma$~significance of
the signal is to minimize systematic effects on the calculation of
the correlation coefficient. We do not find any clear correlation
between the $HR$ and the flux level. Thus we do not find any change
in the spectral behavior in nights where \lsi\ is detected at modest
significance.

The spectral studies on \lsi\ exhibits that the spectrum is soft
(compared to other galactic sources) during all phases in which
\lsi\ is detected at TeV energies. While the flux level changes on
timescales of days and reaches a maximum flux (detected above
$3\sigma$) of $15.6\times 10^{-12}$~cm$^{-2}$s$^{-1}$ the source
shows a spectral photon index of $2.6\pm0.2$, compatible with being
constant.

\section{Timing analysis}

\subsection{Search for intra-night variability}

\begin{figure}[tbp]
  \centering
  \includegraphics[width=\linewidth]{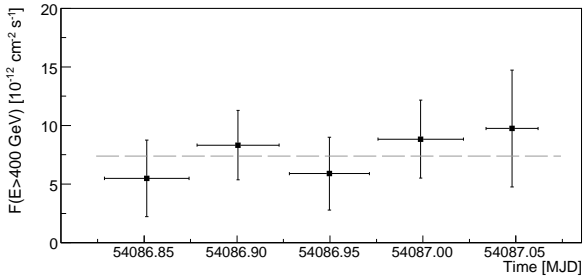}
  \caption{Intra-night integral flux behavior for the longest observed
    night 18$^{\textrm{th}}$ December 2006,
    MJD 54086.95 and orbital phase 0.61. The
    light curve is fitted to a constant flux level with a probability of
    90\% ($\chi^2/\textrm{dof}=1.08/4$). }
  \label{fig:intranight_lc}
\end{figure}

\lsi\ was found to be variable on timescales of days in the previous
observational campaign by MAGIC~\citep{Albert:2006vk}. A still open
question is whether \lsi\ also shows variability on shorter time
scales. We investigated the data for all nights with a significant
flux level ($F(E>400 \mathrm{GeV})>4\times
10^{-12}$~cm$^{-2}$s$^{-1}$) with respect to intra-night variability
on time scales ranging from 30 to 75 minutes, in steps of 15
minutes. This yields 16 suitable nights. Among those, the longest
one was MJD 54086.95 (phase 0.61). Its intranight light curve is
shown in Fig.~\ref{fig:intranight_lc}, where each bin has
$\sim1$~hour width. The light curve is fitted with a constant flux
level with probability of 90\% ($\chi^2/\textrm{dof}=1.08/4$). For
the rest of the nights and tested time scales, the post-trial
probabilities of being chance fluctuations of a constant flux are
all above $32\%$.

We conclude that the VHE fluxes are constant on timescales of $30\,
-\,75$ minutes within the MAGIC sensitivity.

\subsection{Search for periodicity}

The emission of the binary system changes periodically in radio,
optical and X-rays, and the modulation is associated with the orbital
period of the binary system. At higher energies, EGRET measurements
showed hints for variable $\gamma$-ray emission~\citep{tavani:1998},
although no periodicity could be established with these data. One of
the aims of the MAGIC long observational campaign on \lsi\ was to
search for periodic VHE $\gamma$-ray emission.

In order to maximize the sensitivity and accuracy of the timing
analysis we used the data presented in this work together with the
data taken in the first campaign~\citep{Albert:2006vk}, with
observation time of 54 h and covering 6 orbital cycles.

The periodicity analysis was carried out using the formalism
developed by Lomb and Scargle~\citep{Lomb:1976, Scargle:1982}.
This formalism allows to analyze unevenly sampled data while still
keeping the simple exponential probability density function (PDF,
$P(z>z_{0}) = e^{-z_{0}}$)
for Gaussian White Noise (GWN) as valid for the classical Fourier
analysis of evenly sampled data. A remaining problem caused by
the uneven data sampling is that the independent Fourier spacing is
broken, i.e. even a single frequency component can result in a complex
power spectrum with a large number of aliasing peaks.
An other problem is that the mean and variance values that enter the
\lomb\ periodogram have to be estimated from the data themselves.

A practical method to determine the chance probabilities is the
following (see e.g.~\cite{Frescura:2007}):
\begin{enumerate}
\item  A large number of random data series is constructed with a
  Monte Carlo simulation of random fluxes while keeping the sampling
  times fixed.
\item For each random series, we construct a periodogram, sampling it
  for a pre-selected group of frequencies.
\item For each frequency, we compare the periodogram derived from the
  real data set with the probability density function (PDF) obtained
  from the simulated random series, in order to empirically determine
  the (pre-trial) chance probability.
\item The overall (post-trial) chance probability is computed according
  to the following generalization: for each simulated data series we
  inspect the corresponding periodogram, identify the highest Fourier
  power that occurs at any of the pre-selected frequencies, and use
  this value to construct the post-trial PDF.
  It should be noted that this constructed PDF is based on the null
  hypothesis of GWN.
    \end{enumerate}
Integration of the post-trial PDF gives the complementary Cumulative
Distribution Function (cCDF) which is used to determine the
(post-trial) chance probability for a given Fourier power value.

\begin{figure}[tbp]
  \centering
  \includegraphics[width=\linewidth]{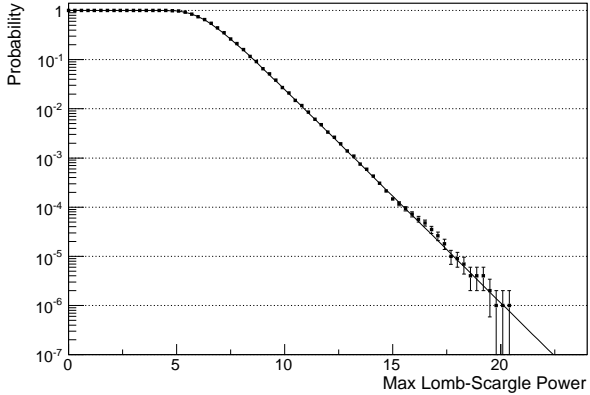}
  \caption{Post-trial complementary Cumulative Distribution Function for the
    \lomb\ power derived from $10^{6}$ random time series.
    The expected cCDF is also indicated (solid line). This function is
    used to estimate the chance probability for powers above 20.}
  \label{fig:LS_density_function}
\end{figure}

In Fig.~\ref{fig:LS_density_function} we show the empirical
post-trial cCDF of the \lomb\ power, estimated via Monte Carlo
simulation of random fluxes. The expected cCDF above a spectral peak
$z_0$ is $F(z>z_0) = 1 - (1- e^{-z_0} )^M$, where $M$ is the number
of independent frequencies. By fitting the PDF for \lsi, we obtain a
probability of 75\% ($\chi^2/\textrm{dof}=263.9/279$) and a number of
independent frequencies of $M=550.8\pm0.6$. This result is used to
estimate the chance probability of the \lomb\ powers.

\begin{figure}[tbp]
  \centering
  \includegraphics[width=\linewidth]{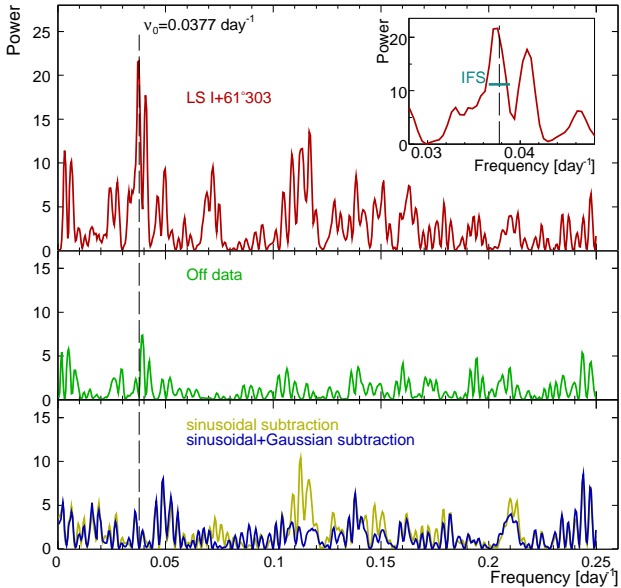}
  \caption{\lomb\ periodogram over the combined
    2005 and 2006 campaigns of \lsi\ data (upper panel) and simultaneous
    background data (middle panel).
    In the lower panel we show the periodograms after subtraction of a
    sinusoidal signal (see Fig.~\ref{fig:sinusfit}) at the orbital
    period (yellow line) and a sinusoidal plus a Gaussian wave form
    (blue line).
    Vertical dashed line corresponds to the orbital frequency.
    Inset: zoom around the highest peak, which corresponds to the
    orbital frequency ($0.0377$d$^{-1}$).
    Its post-trial probability is nearly $10^{-7}$ (see
    Fig.~\ref{fig:LS_density_function}). The IFS is also shown.}
  \label{fig:LS_periodogram}
\end{figure}

In Fig.~\ref{fig:LS_periodogram} (middle panel) we show the
\lomb\ periodogram for an almost (up to detector related effects)
independent background sample, obtained simultaneously with the \lsi\
data (see below for the time intervals).
The highest obtained power is 7.5, which yields a probability of
$0.3$. Thus we obtain no significant probability peaks for any of the
scanned frequencies.

We apply the \lomb\ test to the \lsi\ data and obtain the periodogram
shown in Fig.~\ref{fig:LS_periodogram} (upper panel).
The periodogram is performed with the \lsi\ integral flux above 400
GeV, measured in a time interval $[t_i-\frac{\Delta t}{2},
  t_i+\frac{\Delta t}{2}]$, for $\Delta t=15$~minutes and $i=0,\dots 717$
data points. The overall time range is  $442$ days, which yields an
independent Fourier spacing (IFS) of $\nu_{IFS}=1/T=0.0023$~d$^{-1}$.
We scanned the frequency range $0.0023$-$0.25$~d$^{-1}$ with a
an oversampling factor of 5.

A maximum peak in the \lomb\ periodogram is clearly seen
at frequency $\nu=0.0373$d$^{-1}$, for which we obtain a power of
$\sim22$,  corresponding to a post-trial chance probability of
$2\times10^{-7}$.

Several less prominent but significant peaks are also detected for other
frequencies (e.g. 0.041 d$^{-1}$ with probability $\leq 10^{-5}$).
Those peaks are related to the signal, since they are not
present in the contemporaneous background sample
(Fig.~\ref{fig:LS_periodogram}, middle panel). These are aliasing
peaks of the orbital period of \lsi\ caused by the various gaps in the
data set.

The observational bias due to the moon cycle cannot be the responsible
for the observed peak since this period should otherwise be also
present in the background periodogram.

\begin{figure}[tbp]
  \centering
  \includegraphics[width=\linewidth]{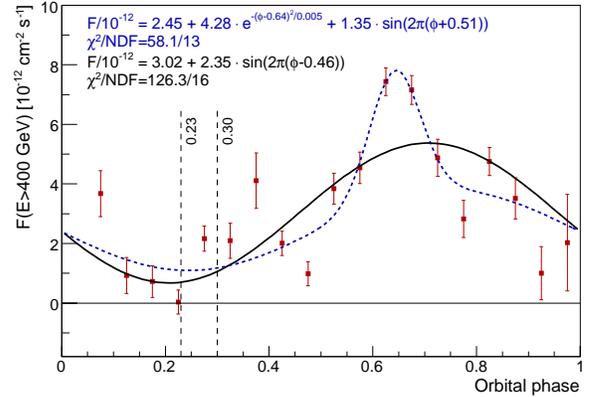}
  \caption{\lsi\ $\gamma$-ray flux above 400~GeV
    obtained from the first and second campaigns, folded
    with the orbital frequency in bins of 0.05 in phase.
    The black curve is a fit to a sinusoidal
    signal. We also fitted a sinusoidal signal plus a Gaussian component
    (blue dotted line), which adjusts better to the data
    (fit parameters are given in the inset). The vertical dashed lines mark the
    two measurements of the periastron passage (according
    to~\cite{Casares:2005wn} and \cite{Grundstrom:2006}).}
  \label{fig:sinusfit}
\end{figure}

The data folded with the peak frequency ($\nu=0.0377$d$^{-1}$) is
presented in Fig.~\ref{fig:sinusfit}, where a sinusoidal fit is
performed ($\chi^2/\textrm{dof}=123.6/16$). Subtracting the obtained
sinusoid from the data, we produce the periodogram shown in
Fig.~\ref{fig:LS_periodogram} (lower panel, yellow line). The peak
associated to the orbital frequency has been removed as expected.
Also the satellite peaks are reduced, but the fact that some of the
other peaks do not achieve a level consistent with the background
test indicates that the signal in the \lsi\ data is not purely
sinusoidal.

To reduce these remaining powers, we fitted the data set with a more
complex signal. Motivated by the data shape, we fitted the data set
with a sinusoidal function plus a Gaussian signal contribution
($\chi^2/\textrm{dof}=58.1/13$), as shown in Fig.~\ref{fig:sinusfit} (blue
dotted line). The corresponding periodogram subtracting this
function to the \lsi\ data set is given in
Fig.~\ref{fig:LS_periodogram} (lower panel, blue line). The orbital
frequency peak has been removed and some of the periodogram peaks
are much more reduced than in the purely sinusoidal subtraction,
giving a better agreement with the background periodogram level.

We performed a Monte Carlo simulation to evaluate the error in the
frequency estimation without any signal shape assumption: we
simulate light curves where the number of $\gamma$-ray and background
candidates are selected randomly from Poisson distributions with a mean
equal to the actually measured distributions of events, arriving in
every given time interval.
The periodogram is calculated for $10^3$ of those randomly generated
series, and the distribution of the resulting peak power frequencies
is fitted with a Gaussian function, yielding an error of
$0.0003$~d$^{-1}$. An accurate peak frequency determination is done by
scanning more frequencies (increasing the oversampling factor)
around the frequency which has maximum probability in the
periodogram, and is found to be (0.0373$\pm$0.0003)~d$^{-1}$,
corresponding to a period of (26.8$\pm$0.2)~days.

\begin{figure}[tbp]
  \centering
  \includegraphics[width=\linewidth]{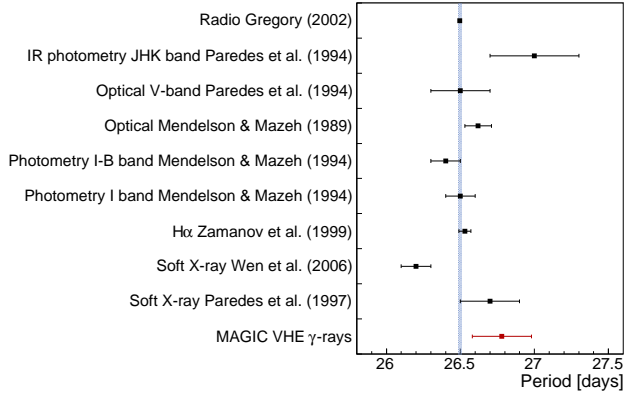}
  \caption{\lsi\ period measurements in different wavelengths. Blue
  band indicates a 3$\sigma$ region around the radio measurement. The
  $\gamma$-ray period (in red) is compatible within 1.5$\sigma$ with
  it.}
  \label{fig:periods_compare}
\end{figure}

In Fig.~\ref{fig:periods_compare} we show the period obtained with
MAGIC data compared to the measurements in other wavelengths. The
most accurate measure of the orbital period is
(26.4960$\pm$0.0028)~days, reported in radio by~\cite{Gregory:2002}.
We also show period measurements reported in near IR and optical
V-band~\citep{Paredes:1994}, optical
wavelengths~\citep{Mendelson:1989}, photometry in the I-B and I
bands~\citep{Mendelson:1994}, H$\alpha$
measurements~\citep{Zamanov:1999}, and soft X-ray measurements
from~\cite{Wen:2006} and \cite{Paredes:1997}. The period obtained
with MAGIC data is compatible with the orbital radio measurement
within 1.5$\sigma$.

\section{Conclusion}
We find that \lsi\ is a periodic $\gamma$-ray binary with an orbital
period of 26.8$\pm$0.2~days (and chance probability $\sim 10^{-7}$),
compatible with the optical, radio and X-ray period. This result
implies that the flux modulation is tied to the orbital period. The
high state in VHE $\gamma$-rays occurs in the same phases as the
X-ray high state. This is especially interesting since we found a
additional hint for X-ray/$\gamma$-ray variability correlation in
the orbital phase 0.85. A strictly simultaneous multi-wavelength
campaign is needed to investigate this correlation in more detail.

We looked for possible intranight variability and found the flux
consistent with being constant within errors in 30-75 minutes
time-scales.

We produce energy spectra for two phase bins $0.5 < \phi < 0.6$ and
$0.6 < \phi < 0.7$ and averaged flux values for several phase bins.
There is clear evidence for a significant change in the VHE
$\gamma$-ray flux level between different phase bins of \lsi. The
spectral photon index does not show this dependence on the phase.
All derived spectral photon indices are compatible with $2.6\pm0.2$
, obtained from the most significant phase bin of \lsi.

We can put constraints to the emission at the periastron passage and
conclude that the system is detected in $\gamma$-rays only in the
phases $0.5-0.9$. Since significant emission is only detected in an
orbital sector off the phases at which the maximum gamma ray flux
should occur under photon-photon absorption (see fig. 5
in~\citealt{2006A&A...451....9D}), the latter can hardly be the only
source of variability in the emission.

Thorough multiwavelength observations will allow us to probe the
intrinsic variability of the non-thermal emission from \lsi\ along
the orbit and can proof possible correlations between the X-ray and
TeV energy bands. This is a necessary step for understanding the
source nature, and the physics underlying the VHE radiation.

\section*{Acknowledgements}

We thank the IAC for the excellent working conditions at the ORM.
The support of the German BMBF, MPG and the YIP of the Helmholtz
Gemeinschaft, the Italian INFN, the Spanish MEC, the ETH Research
Grant TH 34/04 3 and the Polish MNiI Grant 1P03D01028 is gratefully
acknowledged.


\bibliographystyle{astron}
\bibliography{biblio}

\begin{thebibliography}{}

\bibitem[\protect\astroncite{Acciari et~al.}{2008}]{Veritas_lsi}
Acciari, V.~A. et~al.: 2008,
\newblock {\em \apj} {\bf 679}, 1427

\bibitem[\protect\astroncite{Albert et~al.}{2006}]{Albert:2006vk}
Albert, J. et~al.: 2006,
\newblock {\em Science} {\bf 312}, 1771

\bibitem[\protect\astroncite{Albert et~al.}{2007}]{magic:unfolding}
Albert, J. et~al.: 2007,
\newblock {\em Nucl. Instrum. Meth.} {\bf A583}, 494

\bibitem[\protect\astroncite{Albert et~al.}{2008a}]{magic:RF}
Albert, J. et~al.: 2008a,
\newblock {\em Nucl. Instrum. Meth.} {\bf A588}, 424

\bibitem[\protect\astroncite{Albert et~al.}{2008b}]{magiclsi.2006MW}
Albert, J. et~al.: 2008b,
\newblock {\em \apj},
\newblock in press (astro-ph/0801.3150)

\bibitem[\protect\astroncite{Albert et~al.}{2008c}]{crab:2008}
Albert, J. et~al.: 2008c,
\newblock {\em Astroph. J.} {\bf 674}, 1037

\bibitem[\protect\astroncite{{Bednarek}}{2006}]{Bednarek:2006}
{Bednarek}, W.: 2006,
\newblock {\em \mnras} {\bf 371}, 1737

\bibitem[\protect\astroncite{Bock et~al.}{2004}]{Bock:2004td}
Bock, R.~K. et~al.: 2004,
\newblock {\em Nucl. Instrum. Meth.} {\bf A516}, 511

\bibitem[\protect\astroncite{{Bosch-Ramon} et~al.}{2006}]{Bosch-Ramon:2006}
{Bosch-Ramon}, V., {Paredes}, J.~M., {Romero}, G.~E., and {Rib{\'o}}, M.: 2006,
\newblock {\em \aap} {\bf 459}, L25

\bibitem[\protect\astroncite{Casares et~al.}{2005}]{Casares:2005wn}
Casares, J., Ribas, I., Paredes, J.~M., Marti, J., and Allende~Prieto, C.:
  2005,
\newblock {\em \mnras} {\bf 360}, 1091

\bibitem[\protect\astroncite{{Chernyakova} et~al.}{2006}]{2006MNRAS.372.1585C}
{Chernyakova}, M., {Neronov}, A., and {Walter}, R.: 2006,
\newblock {\em \mnras} {\bf 372}, 1585

\bibitem[\protect\astroncite{{Dhawan} et~al.}{2006}]{2006smqw.confE..52D}
{Dhawan}, V., {Mioduszewski}, A., and {Rupen}, M.: 2006,
\newblock in {\em Proceedings of the VI Microquasar Workshop: Microquasars and
  Beyond. September 18-22, 2006, Como, Italy., p.52.1}

\bibitem[\protect\astroncite{{Dubus}}{2006}]{2006A&A...451....9D}
{Dubus}, G.: 2006,
\newblock {\em \aap} {\bf 451}, 9

\bibitem[\protect\astroncite{{Esposito} et~al.}{2007}]{2007A&A...474..575E}
{Esposito}, P., {Caraveo}, P.~A., {Pellizzoni}, A., {de Luca}, A., {Gehrels},
  N., and {Marelli}, M.~A.: 2007,
\newblock {\em A\&A} {\bf 474}, 575

\bibitem[\protect\astroncite{Fegan}{1997}]{Fegan:1997db}
Fegan, D.~J.: 1997,
\newblock {\em J. Phys.} {\bf G23}, 1013

\bibitem[\protect\astroncite{Fomin et~al.}{1994}]{Fomin:1994aj}
Fomin, V.~P. et~al.: 1994,
\newblock {\em Astropart. Phys.} {\bf 2}, 137

\bibitem[\protect\astroncite{Frescura et~al.}{2007}]{Frescura:2007}
Frescura, F.~A.~M., Engelbrecht, C.~A., and Frank, B.~S.: 2007,
\newblock {\em \apj},
\newblock submitted (astro-ph/0706.2225)

\bibitem[\protect\astroncite{Gregory}{2002}]{Gregory:2002}
Gregory, P.~C.: 2002,
\newblock {\em \apj} {\bf 575}, 427

\bibitem[\protect\astroncite{Grundstrom et~al.}{2007}]{Grundstrom:2006}
Grundstrom, E.~D. et~al.: 2007,
\newblock {\em Astrophys. J.} {\bf 656}, 437

\bibitem[\protect\astroncite{Hillas}{1985}]{Hillas_parameters}
Hillas, A.~M.: 1985,
\newblock {\em Proc. 19th ICRC} {\bf 3}, 445

\bibitem[\protect\astroncite{Kniffen et~al.}{1997}]{kniffen}
Kniffen, D.~A. et~al.: 1997,
\newblock {\em Astrophys. J.} pp 126--131

\bibitem[\protect\astroncite{Lomb}{1976}]{Lomb:1976}
Lomb, N.~R.: 1976,
\newblock {\em \apss} {\bf 39}, 447

\bibitem[\protect\astroncite{Maraschi and Treves}{1981}]{Maraschi:1981}
Maraschi, L. and Treves, A.: 1981,
\newblock {\em \mnras} {\bf 194}, 1

\bibitem[\protect\astroncite{Marti and Paredes}{1995}]{Paredes:1995}
Marti, J. and Paredes, J.~M.: 1995,
\newblock {\em \aap} {\bf 298}, 151

\bibitem[\protect\astroncite{Massi et~al.}{2004}]{Massi:2004}
Massi, M. et~al.: 2004,
\newblock {\em \aap} {\bf 414}, L1

\bibitem[\protect\astroncite{Mendelson and Mazeh}{1989}]{Mendelson:1989}
Mendelson, H. and Mazeh, T.: 1989,
\newblock {\em \mnras} {\bf 239}, 733

\bibitem[\protect\astroncite{Mendelson and Mazeh}{1994}]{Mendelson:1994}
Mendelson, H. and Mazeh, T.: 1994,
\newblock {\em \mnras} {\bf 267}, 1

\bibitem[\protect\astroncite{Paredes et~al.}{1994}]{Paredes:1994}
Paredes, J.~M. et~al.: 1994,
\newblock {\em \aap} {\bf 288}, 519

\bibitem[\protect\astroncite{Paredes et~al.}{1997}]{Paredes:1997}
Paredes, J.~M., Marti, J., Peracaula, M., and Ribo, M.: 1997,
\newblock {\em \aap} {\bf 320}, L25

\bibitem[\protect\astroncite{Rolke et~al.}{2005}]{Rolke:2004mj}
Rolke, W., Lopez, A., and Conrad, J.: 2005,
\newblock {\em Nucl. Instrum. Meth.} {\bf A551}, 493

\bibitem[\protect\astroncite{Scargle}{1982}]{Scargle:1982}
Scargle, J.~D.: 1982,
\newblock {\em \apj} {\bf 263}, 835

\bibitem[\protect\astroncite{Tavani et~al.}{1998}]{tavani:1998}
Tavani, M. et~al.: 1998,
\newblock {\em Astrophys. J.} {\bf 648}, L89

\bibitem[\protect\astroncite{Taylor and Gregory}{1982}]{1982ApJ...255..210T}
Taylor, A.~R. and Gregory, P.~C.: 1982,
\newblock {\em \apj} {\bf 255}, 210

\bibitem[\protect\astroncite{Wen et~al.}{2006}]{Wen:2006}
Wen, L., Levine, A.~M., Corbet, R.~H.~D., and Bradt, H.~V.: 2006,
\newblock {\em \apjs} {\bf 163}, 372

\bibitem[\protect\astroncite{Zamanov et~al.}{1999}]{Zamanov:1999}
Zamanov, R.~K., Mart{\'{\i}}, J., Paredes, J.~M., Fabregat, J., Rib{\'o}, M.,
  and Tarasov, A.~E.: 1999,
\newblock {\em \aap} {\bf 351}, 543

\end{thebibliography}

\end{document}